\documentclass[sn-mathphys]{sn-jnl}

\jyear{2021}%

\theoremstyle{thmstyleone}%
\theoremstyle{thmstyletwo}%

\theoremstyle{thmstylethree}%
\numberwithin{equation}{section}
\usepackage[utf8]{inputenc}
\usepackage{amsfonts, latexsym, amsmath, amssymb, fancyhdr, amsthm, graphicx, textcomp}
\usepackage{enumitem}
\usepackage{graphicx}
\usepackage{subfigure}
\usepackage{epstopdf}
\usepackage{epsfig}

\usepackage{url}

\usepackage[english]{babel}
 \usepackage{xcolor}

\usepackage{comment}

\begin{document}

\title[Incorporating separable static heterogeneity]{A systematic procedure for incorporating separable static heterogeneity into compartmental epidemic models}


\author*[1]{\fnm{Odo} \sur{Diekmann}}\email{O.Diekmann@uu.nl}

\author[2]{\fnm{Hisashi} \sur{Inaba}}\email{inaba@ms.u-tokyo.ac.jp}


\affil*[1]{\orgdiv{Mathematical Institute}, \orgname{Utrecht University}, \orgaddress{\street{P.O.Box	80.010}, \city{Utrecht}, \postcode{NL3508TA}, \state{} \country{The Netherlands}}}

\affil[2]{\orgdiv{Graduate School of Mathematical Sciences}, \orgname{The University of Tokyo}, \orgaddress{\street{Komaba 3-8-1}, \city{Meguro-ku, Tokyo}, \postcode{153-8914}, \state{} \country{Japan}}}



\abstract{ }
In this paper, we show how to modify a compartmental epidemic model, without changing the dimension, such that separable static heterogeneity is taken into account. The derivation is based on the Kermack-McKendrick renewal equation.




\maketitle


\begin{center}
\large{In remembrance of Fred Brauer}
\end{center}

\section{Introduction}\label{sec1}

   Up to high age, Fred Brauer has been very active in the field of Mathematical Epidemiology. His books \cite{brauer2012, brauer2019}, lecture notes \cite{brauer2008} and many papers constitute a valuable 
   heritage.  In his paper \cite{brauer2005}, Fred recognizes that Kermack and McKendrick (KM) introduced in \cite{kermack1927} an age-of-infection model that is mathematically represented by a Renewal Equation (RE). Only for very special kernels does the RE reduce to a finite system of ODEs. Or, in other words, compartmental models form a rather restricted subclass of the general KM model. The paper \cite{diekmann2018} provides necessary and sufficient conditions for when a delay equation (i.e., a Delay Differential Equation (DDE) or a RE) allows a finite dimensional reduction. Most of that paper concentrates on linear equations, but Section 9.3 is devoted to the nonlinear KM model.

   In \cite{diekmann2012} and in work in progress \cite{bootsma2022}, we formulate the abstract RE that incorporates static heterogeneity of the host population into the general KM model, see Section 5 below. From a modeling point of view, everything is straightforward. But when it comes to analysing the equation, the infinite dimensional character is a great stumbling block. In Section 8.4 of \cite{diekmann2012}, it is shown that under the assumption of separable mixing, (a slightly less restrictive version of \eqref{5.3} below), we are back to scalar quantities. This facilitates both the computational and the analytical aspects tremendously, see for instance \cite{tkachenko2021, neipel2020, novozhilov2008}. 

   The interpretation of the separability condition can be informally described as follows: whenever the trait/type at the moment of becoming infected is following an a priori given distribution (in particular independently of the trait/type of the infecting individual), newly infected individuals are identical in a stochastic sense and therefore we know how to take averages. From a mathematical point of view, the key point is that various operators have a one-dimensional range.

   After the reduction of the abstract RE to a scalar RE, we can further reduce to an ODE system, provided the time-since-infection kernel has the required form. We claim that this seemingly circuitous derivation of compartmental models, that incorporate separable static host heterogeneity, is, due to its systematic character, far more powerful and efficient than a direct approach that starts from the compartmental model describing a homogeneous host population and then adds heterogeneity. Stated differently: the RE formulation is much more amenable to generalization than the ODE formulation. The aim of the present paper is to establish this systematic procedure and to demonstrate its effectiveness.

    We restrict to models of an outbreak in a closed population, i.e., we ignore both demographic turnover and loss of immunity. This is quite essential. Indeed, we shall first reconsider the homogeneous KM model and reduce it somewhat differently from the manner described in Section 9.3 of \cite{diekmann2018}. This new reduction involves an integration step that only works in the outbreak situation. And it is this new reduction that easily generalizes to the heterogeneous KM model.

   The organization of the paper is as follows. In Section \ref{KM}, we introduce the original KM model, characterized by a kernel $A(\tau)$, with $\tau$ the time elapsed since infection and $A$ the {\it expected} contribution to the force-of-infection. We derive, as in \cite{breda2012}, a RE for the cumulative force-of-infection. In Section \ref{special}, we focus on the special case of a kernel $A$ which is a matrix exponential sandwiched between two vectors. For this special case we deduce the corresponding ODE compartmental system. In Section \ref{two}, we explain how the form of the compartmental model derived in Section \ref{special} relates to the standard form. As concrete examples, we consider the elementary SIR and SEIR models (the treatment of more complicated  examples is postponed till Section \ref{example}).  In Section \ref{hetero}, we turn to heterogeneity. Individuals are characterized by a trait $x$ taking values in a measurable space $\Omega$. The kernel is now a function of three variables, $\tau$, $x$ and $\xi$, with $\tau$ the time elapsed since an individual with trait $\xi$ became infected and $A$ being the expected contribution to the force-of-infection on individuals with trait $x$. Assuming that $A$ is the product of functions $a(x)$, $b(\tau)$ and $c(\xi)$, we derive a scalar renewal equation for the function $w$ such that the cumulative force-of-infection on individuals with trait $x$ equals $a(x) w(t)$. Next we assume that $b$ is a matrix exponential sandwiched between two vectors and reduce to an ODE system. This {\it heterogeneous} ODE system only differs from the corresponding {\it homogeneous} ODE system in the definition of a function $\Psi : \mathbb R \to \mathbb R$. The definition of $\Psi$ involves the functions $a$, $c$ and the measure $\Phi$ describing the distribution of the trait in the host population. The upshot is that one can incorporate separable static heterogeneity into compartmental models by appropriately choosing $\Psi$.
   Section \ref{standard} is devoted to taking heterogeneity into account in the standard form of a compartmental model.  
   In Section \ref{gamma}, inspired by \cite{gomes2022, montalban2022, neipel2020, novozhilov2008, novozhilov2012, tkachenko2021}, 
   we elaborate the special case that $\Omega = (0, \infty)$, $\Phi$ is the Gamma Distribution, $a(x)=x$ and $c(\xi)$ is either identically equal to one or equal to $\xi$.  Section \ref{example} is devoted to examples and in the final Section \ref{final} we collect some concluding remarks.

\section{The general Kermack-McKendrick model}\label{KM}

Let $S$ be the size of the subpopulation of susceptible individuals, and let $F$ denote the force-of-infection. In a closed population, the incidence is equal to both $F S$ and to $- dS/dt$, so we have
\begin{equation}\label{2.1}
\frac{dS}{dt} = - F S.
\end{equation}

The essence of the KM model is the constitutive equation that expresses $F$ in terms of contributions by individuals that became infected before the current time:
\begin{equation}\label{2.2}
 F(t) = \int_{0}^{\infty}A(\tau) F(t-\tau)S(t-\tau) d\tau.
\end{equation}

Here the one-and-only (apart from $N$, the total host population size) model ingredient $A$ describes the expected contribution to the force-of-infection as a function of the time $\tau$ elapsed since infection took place. In this top-down approach we postpone a specification of the stochastic processes that underly the word {\it expected} (in general, these concern both within host processes, in particular the struggle between the pathogen and the immune system, and the between host contact process). Indeed, KM wanted to know what general conclusions can be drawn {\it without} providing such a specification.

   Now imagine that $F$ was negligible in the infinite past. We introduce the cumulative force-of-infection $w$ defined by
\begin{equation}\label{2.3}
  w(t) := \int_{-\infty}^{t} F(\sigma) d\sigma.
  \end{equation}

Suppose that $S(-\infty)=N$. Then integration of \eqref{2.1} yields

\begin{equation}
  S(t) = N e^{-w(t)}.
  \end{equation}
So note, incidentally, that one can equivalently characterize $w$ by the relation $w = - \log(S/N)$.

Integration of \eqref{2.2} with respect to time over $(-\infty,t]$ leads, upon replacing $FS$ by $-S'$ and a change of the order of the integrals, to the RE
\begin{equation}\label{2.5}
  w(t) = \int_{0}^{\infty} A(\tau) \Psi(w(t - \tau)) d\tau,
  \end{equation}
where 
\begin{equation}\label{2.6}
  \Psi(w) := N ( 1 - e^{-w}),
\end{equation}
which corresponds to the subpopulation of no longer susceptible individuals, given the cumulative force of infection.

As a side remark, we mention that \eqref{2.5} can be considered as a deterministic version of the Sellke construction, as described in, for instance, Section 3.5.2 of \cite{diekmann2012}.

\section{The special case in which reduction to a compartmental model is possible}\label{special}

Suppose there exist an integer $n$, $1 \times n$, $n \times 1$-matrices $U$, $V$ and an $n \times n$-matrix $\Sigma$ such that
\begin{equation}\label{3.1}
  A(\tau) = U e^{\tau \Sigma} V,
\end{equation}
then, in a sense, we have a state representation for the (one-time) input-(distributed-time) output map $A$. The matrix $\Sigma$ generates the autonomous Markov chain dynamics of the state, the scalar quantity {\it incidence} is an input along the fixed vector $V$ and the $i$-th component of the vector $U$ measures the output, as a contribution to the force-of-infection, of the $i$-th state. System \eqref{4.1} below and the concrete examples \eqref{4.5} and \eqref{4.7} should clarify this somewhat vague description.

   The claim is that the RE \eqref{2.5} reduces to an ODE system when \eqref{3.1} holds. To substantiate the claim, we define the $n$-vector valued function $Q$ of time by
\begin{equation}\label{3.2}
 \begin{aligned}
  Q(t) :&= \int_{0}^{\infty} e^{\tau \Sigma} V \Psi(w(t-\tau)) d\tau \cr
  &= \int_{-\infty}^{t} e^{(t-\sigma) \Sigma} V \Psi(w(\sigma)) d\sigma.
\end{aligned}
\end{equation}
It follows that
\begin{equation}\label{3.3}
\frac{dQ}{dt} = \Sigma  Q + V \Psi(w).
\end{equation}
On the other hand, it follows from \eqref{3.1}, \eqref{3.2} and \eqref{2.5} that

\begin{equation}\label{3.4}
  w(t) = U Q(t).
\end{equation}

 Combining \eqref{3.3} and \eqref{3.4} we obtain the closed system of ODE
\begin{equation}\label{3.5}
\frac{dQ}{dt} = \Sigma  Q + V \Psi(U Q).
\end{equation}

Note that, conversely, given a solution of \eqref{3.5}, we can define $w$ by \eqref{3.4} and verify that $w$ satisfies \eqref{2.5}. For reasons explained in the next section, we call \eqref{3.5} the {\it integrated form} of the compartmental model corresponding to $\Sigma$, $V$ and $U$.

   For completeness, we like to mention that the assumption \eqref{3.1} immediately allows us to calculate basic indices for the KM model as follows:
\begin{enumerate}
\item	the basic reproduction number is given by  
\begin{equation}\label{3.6}
R_0 = N\int_{0}^{\infty}A(\tau)d\tau=- N U \Sigma^{-1} V.
\end{equation}
\item	the Euler-Lotka equation is 
\begin{equation}\label{3.7}
1 = N\int_{0}^{\infty}e^{-\lambda \tau}A(\tau)d\tau=N U (\lambda I - \Sigma)^{-1} V,
\end{equation} 
and the intrinsic growth rate $r$ is given by its real root. 

\item the generation time, here denoted by $T$,  is given by
\begin{equation}\label{3.8}
T:=\frac{\int_{0}^{\infty}\tau A(\tau) d\tau}{\int_{0}^{\infty}A(\tau)d\tau} =-\frac{U \Sigma^{-2} V}{  U \Sigma^{-1} V}.
\end{equation}
\end{enumerate}

\section{Two ways of formulating compartmental models}\label{two}

We start from \eqref{2.1}-\eqref{2.2}, make assumption \eqref{3.1}, and introduce
\begin{equation}\label{4.1}
       Y(t) := \int_{0}^{\infty} e^{\tau \Sigma}  V F(t-\tau) S(t-\tau) d\tau,
\end{equation}
to {\it count} the individuals, that were infected before time $t$, on the basis of their state at time $t$. 
Then we can deduce, as detailed in Section 9.3 of \cite{diekmann2018}, the ODE system
\begin{equation}\label{4.2}
\begin{aligned}
&\frac{dS}{dt} = - F S,\cr
&\frac{dY}{dt} = \Sigma Y +  (F S)V,
\end{aligned}
\end{equation}
with
\begin{equation}\label{4.3}
       F := U Y.
\end{equation}
There are good reasons to call this the {\it standard form} of the compartmental model corresponding to $\Sigma$, $V$ and $U$. 

We claim that $Q$, as introduced in Section 3, is the integral of $Y$, i.e., 
\begin{equation}\label{4.4}
    Q(t) = \int_{-\infty}^{t} Y(\sigma) d\sigma.
    \end{equation}
In fact, from \eqref{4.1} and \eqref{4.4}, we obtain
\begin{equation}\label{4.5}
\begin{aligned}
Q(t)&= \int_{-\infty}^{t} Y(\sigma) d\sigma=\int_{-\infty}^{t}d\sigma \int_{0}^{\infty} e^{\tau \Sigma}  V (-\dot{S}(\sigma-\tau)) d\tau \cr
&= \int_{0}^{\infty} e^{\tau \Sigma}  V (N-S(t-\tau)) d\tau \cr
&= \int_{0}^{\infty} e^{\tau \Sigma}  V \Psi(t-\tau) d\tau= \int_{-\infty}^{t} e^{(t-\tau) \Sigma}  V \Psi(w(\tau)) d\tau,
\end{aligned}
\end{equation}
where we used \eqref{2.6}.  Note that convergence of the integral in \eqref{4.4} requires that $Y$ converges to zero at $-\infty$. 
 Differentiating the above equation \eqref{4.5} with respect to $t$, we recover \eqref{3.3}, i.e.,
\begin{equation}\label{4.6}
\frac{dQ}{dt} = \Sigma  Q + V \Psi(w).
\end{equation}
Since $w(t)=UQ(t)$, we arrive at the integrated compartmental model \eqref{3.5}.

A minor advantage of the integrated form is that the dimension is $n$, not $n+1$. In the present context, the key favourable feature is that the integrated form extends seamlessly to the separable heterogeneous setting (while the standard form does not).

To illustrate the integrated formalism, we consider the two most basic examples. If we write the standard form of the SIR model as
\begin{equation}\label{4.6}
\begin{aligned}
&\frac{dS}{dt} = - \beta I S, \cr
&\frac{dI}{dt} =  -\alpha I + \beta I S, 
\end{aligned}
\end{equation}
the integrated form reads
\begin{equation}\label{4.7}
\frac{dQ}{dt} = -\alpha Q + \Psi(\beta Q),
\end{equation}
where 
\[Q(t)=\int_{-\infty}^{t}I(\sigma)d\sigma,\]
 and $\Psi$ is defined in \eqref{2.6}.  So here $n=1$, $V=1$, $U=\beta$ and $\Sigma=-\alpha$.

If we write the standard form of the SEIR model as
\begin{equation}\label{4.8}
\begin{aligned}
&\frac{dS}{dt} = - \beta I S, \cr
&\frac{dE}{dt} =  - \gamma E + \beta I S, \cr
&\frac{dI}{dt}  =   \gamma E - \alpha I,
\end{aligned}
\end{equation}
the integrated form reads
\begin{equation}\label{4.9}
\begin{aligned}
&\frac{dQ_1}{dt} = - \gamma Q_1         + \Psi(\beta Q_2), \cr
 &\frac{dQ_2}{dt}  =   \gamma Q_1  - \alpha Q_2,
 \end{aligned}
 \end{equation}  
where
\[\begin{pmatrix} Q_1(t) \cr Q_2(t)\end{pmatrix}=\begin{pmatrix}\int_{-\infty}^{t}E(\sigma)d\sigma \cr \int_{-\infty}^{t}I(\sigma)d\sigma \end{pmatrix}.\]
So here $n=2$ and we have 
\begin{equation}\label{4.10}
V =\begin{pmatrix}1 \cr 0 \end{pmatrix},  ~U = \begin{pmatrix}0  & \beta \end{pmatrix}  ,  ~\Sigma= \begin{pmatrix}-\gamma & 0 \cr \gamma & -\alpha \end{pmatrix}.                                        
        \end{equation}

\section{Incorporating heterogeneity}\label{hetero}

 ﻿Let host individuals be characterized by a trait $x$, with $x$ ranging in a measurable space $\Omega$ (the advantage of this somewhat abstract formulation is that $x$ may be a discrete variable, a continuous variable or a mixture of these two possibilities, in the sense that $x$ has both a discrete and a continuous component). The kernel $A$ now has three arguments, $\tau$, $x$ and $\xi$. The variable $x$ specifies the trait of the individual that is at risk of becoming infected, while $\xi$ and $\tau$ specify, respectively, the trait and the time-since-infection of an infected individual. 
 
   We want a formalism that captures both the case where $\Omega$ is discrete (for instance a finite set) and the case where $\Omega$ is continuous. To achieve this, we describe the population composition by a measure $\Phi$ on $\Omega$ and, for any $t$, a bounded measurable function $s(t, \cdot)$ such that of the individuals with trait $x$, a fraction $s(t,x)$ is still susceptible at time $t$ (in other words: $s(t,x)$ is the probability that an individual with trait $x$ is susceptible at time $t$ and $s(-\infty,x)=1$).
 
   We replace \eqref{2.1}-\eqref{2.2} by, respectively
  
\begin{equation}\label{5.1}
\frac{\partial s}{\partial t} (t,x) = - F(t,x) s(t,x),
\end{equation}
and
\begin{equation}\label{5.2}
   F(t,x) =N\int_{0}^{\infty} \int_\Omega  A(\tau, x, \xi) F(t-\tau, \xi) s(t-\tau, \xi) \Phi(d\xi) d\tau.
\end{equation}

When functions $a$, $b$ and $c$ exist such that
\begin{equation}\label{5.3}
   A(\tau, x, \xi) = a(x) b(\tau) c(\xi),
   \end{equation}
we deduce from \eqref{5.2} that $F$ is the product of $a(x)$ and a function of time. This motivates us to introduce a function $w$ such that
\begin{equation}\label{5.4}
\int_{-\infty}^{t} F(\sigma,x) d\sigma = a(x) w(t).
\end{equation}

Next we integrate \eqref{5.2} with respect to time over $(-\infty, t]$, while using \eqref{5.1} to replace the product $F s$ by the time partial derivative of $s$. After dividing out the factor $a(x)$ that occurs at both sides, we obtain
\begin{equation}\label{5.5} 
 w(t) = \int_{0}^{\infty} b(\tau) \Psi(w(t-\tau)) d\tau,
 \end{equation}
with $\Psi$ now defined by
\begin{equation}\label{5.6}
\Psi(w) := N \int_{\Omega}  c(\xi) (1 - e^{- a(\xi) w} ) \Phi(d\xi), 
\end{equation}
with $\Phi$ the measure that describes the probability distribution of the trait in the host population. Note that from \eqref{5.6} we recover the earlier definition \eqref{2.6} in the special (homogeneous) case that both $a$ and $c$ are identically  equal to 1.

 Also note that when $a$ is identically equal to one, \eqref{5.6} just states that we can simply work with the average value of $c$. So, as indeed emphasized in Chapter 2 of \cite{diekmann2012}, heterogeneity of infectiousness is, in the large numbers deterministic limit, simple : just take the expected value. 

    Essentially, \eqref{2.5} and \eqref{5.5} are the same. When $a$ is not constant, \eqref{5.6} differs from \eqref{2.6}.
   When
\begin{equation}\label{5.7}
  	b(\tau) = U e^{\tau \Sigma} V,
  	\end{equation}
we can define $Q$ as before, see \eqref{3.2}, and retrieve both \eqref{3.5} and \eqref{3.4}. We conclude that in terms of the integrated formulation of a compartmental model, we can incorporate separable heterogeneity by simply redefining the function $\Psi$. The fact that \eqref{5.6} involves an integral is of little to no importance for the theory. But when one wants to study \eqref{3.5} numerically, it is a nuissance. As noted before by various authors (cf. \cite{gomes2022, montalban2022, neipel2020, novozhilov2008, novozhilov2012, tkachenko2021}), in a special case one can replace the integral by an explicit expression. In Section \ref{gamma} we shall provide the details.

\section{The standard form : a recipe}\label{standard}

   In Section \ref{example} we present a model for a heterosexually transmitted disease. In that case we have to distinguish between newly infected males and newly infected females. So there are TWO states-at-infection. 
      Here we focus on models in which there is only one state-at-infection, represented by the vector $V$ (but note that $V$ may be a probability vector with several non-zero components, see Example 8.1).
   
   Our starting point is the compartmental system
\begin{equation}\label{6.1}
\begin{aligned}
               &\frac{dS}{dt}  =  - F S, \cr
&  \frac{dY}{dt} =  \Sigma Y + (F S) V,\cr
&          F = U Y.
\end{aligned}
\end{equation}

We want to incorporate separable static heterogeneity as described by the trait space $\Omega$, the trait distribution $\Phi$, the relative trait-specific susceptibility $a$ and the relative trait-specific infectiousness $c$. Here {\it relative} means that we single out a representative $\bar{x}$ in $\Omega$ and normalise $a$ and $c$ by requiring that $a(\bar{x}) = 1$, $c(\bar{x}) = 1$ (note that \eqref{5.3} offers the possibility to thus normalise $a$ and $c$; if $\Omega$ is a continuum, it makes sense to choose the mean trait as $\bar{x}$).

Then $\bar{s}(t):=s(t,\bar{x})$ denotes the fraction of the individuals with the trait $\bar{x}$ that escaped infection up to the current time. For arbitrary trait $x$, the corresponding probability equals 
\begin{equation}\label{6.2}
s(t,x)=\bar{s}(t)^{a(x)},
\end{equation}
because $s(t,x)=\exp(-a(x)w(t))$.
Hence the fraction of the total population that is still susceptible is given by
\begin{equation}\label{6.3}
     s_{{\rm tot}} = \int_\Omega  \bar{s}^{a(\xi)} \Phi(d\xi).
     \end{equation}
     Moreover, since $a(\bar{x})=1$, we have $w(t)=-\log \bar{s}(t)$. Thus knowledge of $\bar{s}$ is sufficient to determine both $s(t,x)$ and $w(t)$.

On the other hand, in analogy with \eqref{4.1}, let us define the trait-specific $y$-variable by
\begin{equation}\label{y}
y(t,\xi):=\int_{0}^\infty e^{\tau \Sigma}VF(t-\tau,\xi)s(t-\tau,\xi)d\tau,
\end{equation}
and let $Y$ now denote the weighted $y$-variable:
\begin{equation}
\begin{aligned}\label{Y}
Y(t)&=N\int_{\Omega}c(\xi)y(t,\xi)\Phi(d\xi) \cr
&=\int_{-\infty}^{t} e^{(t-\sigma) \Sigma}V N\int_{\Omega}c(\xi)F(\sigma,\xi)s(\sigma,\xi)\Phi(d\xi) d\sigma.
\end{aligned}
\end{equation}
Note that, since the dynamics of infected individuals is independent of the trait and linear, it does not matter whether we apply the weight factor $c$ when computing the output by applying $U$ or right at the start, i.e., immediately after the infection took place.

Now we formulate the standard compartmental system which takes into account the static heterogeneity:

\

{\bf Claim}
The {\it heterogeneous} compartmental system consisting of \eqref{3.5} with $\Psi$ defined by \eqref{5.6}, has the standard form representation

\begin{equation}\label{6.4}
\begin{aligned}
&            \frac{d\bar{s}}{dt}  = - \bar{F} \bar{s},\cr
&     \frac{dY}{dt}  =  \Sigma Y  + (\bar{F} \Psi'(- \log \bar{s}) ) V, \cr
&            \bar{F} = U Y,
\end{aligned}
\end{equation}
where $\bar{s}(t)=s(t,\bar{x})$, $\bar{F}(t)=F(t,\bar{x})$ and explicitly we have
\begin{equation}\label{6.5}
    \Psi'(w)  =  N \int_\Omega  c(\xi) a(\xi) e^{- a(\xi) w }  \Phi(d\xi).
    \end{equation}
\begin{proof}
It follows from \eqref{5.2}, \eqref{5.3}, \eqref{y} and \eqref{Y} that
\begin{equation}
\begin{aligned}
\bar{F}(t)&=N\int_{0}^{\infty} \int_\Omega  a(\bar{x})b(\tau)c(\xi) F(t-\tau, \xi) s(t-\tau, \xi) \Phi(d\xi) d\tau \cr
&=\int_{0}^{\infty}Ue^{\tau \Sigma}V N \int_\Omega c(\xi) F(t-\tau, \xi) s(t-\tau, \xi) \Phi(d\xi) d\tau \cr
&=U Y(t).
\end{aligned}
\end{equation}
Differentiation of \eqref{3.5} with respect to time yields
\begin{equation}    \frac{dY}{dt} = \Sigma Y + (\Psi'(U Q) U Y ) V.
\end{equation}
We identify $U Q$ with $- \log \bar{s}$ , since this is what we obtain when we integrate 
\begin{equation}
      \frac{d\bar{s}}{dt} = - U Y \bar{s}.
      \end{equation}
\end{proof}

\section{The Gamma Distribution}\label{gamma}

Let $\Omega = [0, \infty)$ and let $a(x) = x$, i.e., let the trait correspond directly to relative susceptibility. 
Let $\Phi$ be the Gamma Distribution with mean 1 and variance $p^{-1}$ . In other words, let $\Phi$ have density
\begin{equation}\label{7.1}
       x \to   \frac{p^p}{\Gamma(p)}  x^{p-1}  e^{-px}.
       \end{equation}

The key feature is that under these assumptions we can evaluate the integral in \eqref{5.6} when $c$ is a (low order) polynomial and thus obtain an explicit expression for $\Psi$. The underlying reason is that we deal with (a derivative of) the Laplace Transform of $\Phi$, which is itself explicitly  given by
\begin{equation}\label{7.2}
   \hat{\Phi}(\lambda)  = \left(\frac{\lambda}{p} + 1\right)^{-p}.
\end{equation}

If the trait has no influence on infectiousness, i.e., $c$ is identically equal to 1, we have
\begin{equation}\label{7.3}
   \Psi(w) =N\left[ 1 - \left(\frac{w}{p} + 1\right)^{-p} \right],
\end{equation}
while if infectiousness too is proportional to the trait, i.e., $c(\xi)=\xi$, we obtain
\begin{equation}\label{7.4}
   \Psi(w) = N\left[1 - \left(\frac{w}{p} + 1\right)^{-p-1}\right].
   \end{equation}

In \cite{bootsma2022}, we compare and contrast these special cases in terms of $R_0$ , the Herd Immunity Threshold and the final size. Here we limit ourselves to the observation that \eqref{3.5}, with either \eqref{7.3} or \eqref{7.4}, enables modelers to  study rather easily the impact of separable heterogeneity on the dynamics of their favourite compartmental model, cf. \cite{neipel2020}.

Now recall that the standard form involves $\bar{s}$, the value of $s$ in a representative point $\bar{x}$.
In the special case of the Gamma Distribution, we can work with $s_{{\rm tot}}$ instead of $\bar{s}$ and replace \eqref{6.4} by
\begin{equation}\label{7.5}
\begin{aligned}
&           \frac{d s_{{\rm tot}}}{dt}  =  - \bar{F} s_{{\rm tot}}^{1 + \frac{1}{p}} \cr
  &    \frac{dY}{dt}   =   \Sigma Y  +  (\bar{F} H(s_{{\rm tot}}) ) V\cr
&             \bar{F} =   U Y
\end{aligned}
\end{equation}
with
\begin{equation}\label{7.6}
H(s)=\begin{cases}                           N s^{1+\frac{1}{p}} &             {\rm   if} ~c ~{\rm is~ identically~ equal~ to~ 1},\cr
                          N \left(1+\frac{1}{p}\right) s^{1+\frac{2}{p}} &  {\rm  if}~ c(\xi) = \xi.
                          \end{cases}
                          \end{equation}

The derivation of \eqref{7.5} starts by choosing $\bar{x} = 1$ (=  mean trait), $a(\bar{x})=1$ and rewriting \eqref{6.3} in this special case as
\begin{equation}\label{7.7}
   s_{{\rm tot}} = \hat{\Phi}( - \log \bar{s}).
   \end{equation}

Since $\hat{\Phi}$ is given explicitly by \eqref{7.2}, one can express derivatives of $\hat{\Phi}$ in terms of $\hat{\Phi}$ itself. And derivatives correspond to incorporating powers of the variable in the function that is Laplace transformed.
We refer to \cite{novozhilov2008} for an early derivation of \eqref{7.5} and for references to still earlier work.

\section{Some examples}\label{example}

{\begin{flushleft}
{\bf Example 1}
\end{flushleft}

   The diagram depicted in Figure 1 gives a concise representation of the compartmental model that we consider in this section as a slightly more complex example. Our first aim is to illustrate how one obtains the model ingredients $U$, $\Sigma$, $V$ from such a diagram.

   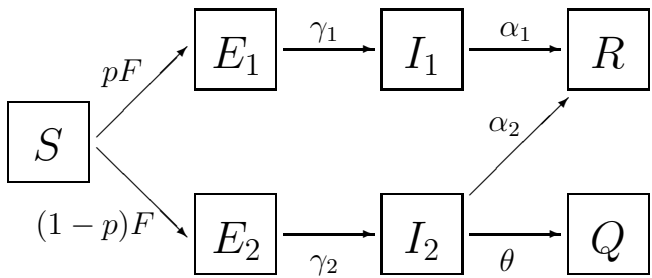
\begin{figure}[h]
\begin{center}
\begin{picture}(270,100)
\put(10,35){\framebox(30,30)}
\put(80,0){\framebox(30,30)}
\put(80,70){\framebox(30,30)}
\put(150,0){\framebox(30,30)}
\put(150,70){\framebox(30,30)}
\put(220,70){\framebox(30,30)}
\put(220,0){\framebox(30,30)}

\put(19,43){\LARGE $S$}
\put(86,78){\LARGE $E_1$}
\put(86,8){\LARGE $E_2$}
\put(158,78){\LARGE $I_1$}
\put(158,8){\LARGE $I_2$}
\put(228,78){\LARGE $R$}
\put(228,8){\LARGE $Q$}

\put(45,73){\large $pF$}
\put(20,15){\large $(1-p)F$}
\put(123,90){\large $\gamma_1$}
\put(123,2){\large $\gamma_2$}
\put(194,90){\large $\alpha_1$}
\put(190,54){\large $\alpha_2$}
\put(194,1){\large $\theta$}

\put(43,52){\vector(1,1){34}}
\put(43,48){\vector(1,-1){34}}
\put(113,85){\vector(1,0){34}}
\put(113,15){\vector(1,0){34}}
\put(183,85){\vector(1,0){34}}
\put(183,15){\vector(1,0){34}}
\put(183,30){\vector(1,1){37}}
\end{picture}
\end{center}
\caption{SEIR model with asymptomatic infection and quarantine}
\end{figure}

   The index 1 denotes asymptomatic individuals, the index 2 symptomatic individuals. As the symptoms get noticed, a diagnosis is possible for symptomatic individuals and subsequently they may be put into quarantaine, i.e., enter $Q$. The two types of individuals occur with ratio $p$ : $1-p$ , with $p$ a parameter. So the state-at-infection/birth is a probability vector $V$ with two non-zero components. Immediately following infection an individual is Exposed but not yet Infectious. The sojourn times of the various compartments are all exponentially distributed with a parameter specified in the diagram (in the form of a name label, so as a parameter, not as a number with dimension 1/time). The compartments $R$ and $Q$ aid the bookkeeping, but their contents is irrelevant for future dynamics, so we do not incorporate them into the population state vector.
   
   As is customarily done, we use the same characters to denote a compartment and the contents of this compartment. Define the 4-vector $Y$ by
\begin{equation}\label{8.1}
Y =\begin{pmatrix}   E_1 \cr     E_2 \cr I_1 \cr I_2 \end{pmatrix},
\end{equation}
the vector $V$ by  
\begin{equation}\label{8,2}
V =  \begin{pmatrix} p \cr        1-p \cr                  0 \cr
                 0
\end{pmatrix},
\end{equation}
the vector $U$  by
\begin{equation}\label{8.3}
U=\begin{pmatrix} 0 &   0 &   \beta_1 &   \beta_2 \end{pmatrix},
\end{equation}
and the matrix $\Sigma$ by
\begin{equation}\label{8.4}
\Sigma =\begin{pmatrix}   -\gamma_1  &            0     &             0    &                0 \cr
                                   0   &           -\gamma_2  &         0     &               0 \cr
                \gamma_1    &        0      &        -\alpha_1    &           0 \cr
                                   0       &         \gamma_2  &        0    &      -(\alpha_2 + \theta)
                                   \end{pmatrix},
                                   \end{equation}
then  \eqref{4.1}-\eqref{4.2} is the standard representation of the compartmental model specified by the diagram. So if we define $Q$ by \eqref{4.4} then we obtain the integrated representation \eqref{3.5}.
 
   Either by a direct consideration, or by verifying that
   \begin{equation}\label{8.5}
-\Sigma^{-1}=\begin{pmatrix}
                              \frac{1}{\gamma_1}&               0 &                             0    &                 0 \cr
                                    0         &        \frac{1}{\gamma_2} &                      0    &                 0\cr
      \frac{1}{\alpha_1} &                  0     &                   \frac{1}{\alpha_1}       &       0\cr
                                    0     &          \frac{1}{\alpha_2 + \theta} &           0      &        \frac{1}{\alpha_2 + \theta}
                                    \end{pmatrix},
                                    \end{equation}
and applying the formula \eqref{3.6}, we obtain
\begin{equation}\label{8.6}
R_0 = N \left\{ p \frac{\beta_1}{\alpha_1}    +   (1-p) \frac{\beta_2}{\alpha_2 + \theta} \right\}.
\end{equation}
Similarly we obtain from \eqref{3.8} that the generation time is given by
\begin{equation}\label{8.7}
T=\frac{ p \frac{\beta_1}{\alpha_1} ( \frac{1}{\gamma_1} + \frac{1}{\alpha_1})     +   (1-p)\frac{\beta_2}{\alpha_2 + \theta} ( \frac{1}{\gamma_2}  +  \frac{1}{\alpha_2 + \theta}) }{ p \frac{\beta_1}{\alpha_1}    +   (1-p) \frac{\beta_2}{\alpha_2 + \theta}  }.
\end{equation}


\

\begin{flushleft}
{\bf Example 1, continued}
\end{flushleft}

Next let us introduce immune system related heterogeneity in the sense that we distinguish between standard individuals, which we label 1, and partly immune individuals, which we label 2. The relative susceptibility of type 2 individuals is given by the parameter $\epsilon_1$ and the relative infectiousness by the parameter $\epsilon_2$. In the present context it does not matter whether the immunity results from an earlier outbreak or from vaccination. But we assume it exists before the outbreak that we model is initiated or, in other words, that it does not result from control measures during the outbreak. 

   Let $N = N_1 + N_2$ with $N_1$ and $N_2$ the size of the subpopulation of individuals of type, respectively, 1 and 2. Then $\Phi$ has components $N_1/N$ and $N_2/N$.
We may choose $1$ as $\bar{x}$ and let $a$ have components 1 and $\epsilon_1$ and let $c$ have components 1 and $\epsilon_2$. 
    It follows that
\begin{equation}\label{8.8}
\Psi(w) =  N_1 (1-e^{-w})  +  N_2 \epsilon_2 (1- e^{- \epsilon_1 w})
\end{equation}

\begin{equation}\label{8.9}
\Psi'(- \log \bar{s}) = N_1 \bar{s} + N_2 \epsilon_1 \epsilon_2 \bar{s}^{\epsilon_1}.
\end{equation}

One can now consider \eqref{3.5} or \eqref{6.4} with the above definitions of $U$, $\Sigma$, $V$ and $\Psi$ (resp. $\Psi'$) and study, for instance, numerically how peak size is influenced by the parameters $\epsilon_1$, $\epsilon_2$ and $N_1$ (for given $N$).

Note that when we choose $\epsilon_1=\epsilon=\epsilon_2$, this example allows for an alternative interpretation: as a result of a control measure, individuals reduce their social activity with a factor $\epsilon$, but only a fraction $N_2/N$ complies, the complementary fraction does not reduce its social activity.

\

\begin{flushleft}
{\bf Example 2}
\end{flushleft}

  In the formulation of our results, we have restricted to the situation in which one vector $V$ and one vector $U$ suffice. In \cite{diekmann2018}, the generic result actually concerns systems of REs for which one needs as many vectors $V$ and $U$ as the number of components of the system. Here we illustrate how this works by concentrating on the epidemiologically relevant example of a heterosexually transmitted disease (or a disease transmitted by a vector).
  
  Let $S_i$ $(i=1,2)$ be the size of the susceptible population, where the index 1 denotes males and the index 2 females (or the host and the vector, respectively). 
We postulate that
\begin{equation}\label{8.10}
\begin{aligned}
&\frac{dS_i}{dt} = - F_i S_i,\cr
&\begin{pmatrix} F_1(t) \cr F_2(t)\end{pmatrix} = \int_{0}^{\infty}A(\tau) \begin{pmatrix}F_1(t-\tau)S_1(t-\tau)\cr F_2(t-\tau)S_2(t-\tau) \end{pmatrix} d\tau,
\end{aligned}
\end{equation}
where
\begin{equation}\label{8.11}
A(\tau):=\begin{pmatrix} 0 & A_{12}(\tau) \cr A_{21}(\tau) & 0 \end{pmatrix}.
\end{equation}

Define the cumulative force of infection as
\begin{equation}\label{8.12}
w(t)=\begin{pmatrix} \int_{-\infty}^{t}F_1(\sigma)d\sigma \cr \int_{-\infty}^{t}F_2(\sigma)d\sigma \end{pmatrix}.
\end{equation}
Then we obtain the renewal equation:
\begin{equation}\label{8.13}
w(t)=\int_{0}^{\infty}A(\tau)\Psi(w(t-\tau))d\tau,
\end{equation}
with
\begin{equation}\label{8.14}
\Psi(w)=\begin{pmatrix} N_1(1-e^{-w_1}) \cr N_2(1-e^{-w_2})\end{pmatrix}, 
\end{equation}
where $N_1$ denotes the total size of the male population and $N_2$ is the total size of the female population.

For the compartmental case, we have
\begin{equation}\label{8.15}
A(\tau):=\begin{pmatrix} 0 & U_2e^{\tau \Sigma_2}V_2 \cr U_1e^{\tau \Sigma_1}V_1 & 0 \end{pmatrix}.
\end{equation}
Here $U_i$, $V_i$ are $1 \times n_i$, $n_i \times 1$ matrices and $\Sigma_i$ is an $n_i \times n_i$ matrix.  It seems to make sense to assume that $n_1=n_2$ and  perhaps $\Sigma_1=\Sigma_2$.  But if within host processes are different for males and females, we should allow for $n_1 \neq n_2$ and $\Sigma_1 \neq \Sigma_2$.
This is anyhow a reasonable assumption for the host-vector situation.

Define vectors $Q_1$ and $Q_2$ by
\[w_1=U_2 Q_2, \quad w_2=U_1 Q_1,\]
then
\begin{equation}\label{8.16}
\begin{aligned}
&\frac{dQ_1}{dt}=\Sigma_1 Q_1+\Psi_1(U_2Q_2)V_1,\cr
&\frac{dQ_2}{dt}=\Sigma_2 Q_2+\Psi_2(U_1Q_1)V_2.\cr
\end{aligned}
\end{equation}
This is the integrated version of the following standard form:
\begin{equation}\label{8.17}
\begin{aligned}
&\frac{dS_i}{dt} = - F_i S_i, & (i=1, 2),\cr
&\frac{dY_i}{dt}=\Sigma_i Y_i+F_iS_iV_i,\cr
&F_1=U_2Y_2, \cr
&F_2=U_1Y_1.
\end{aligned}
\end{equation}

\

\begin{flushleft}
{\bf Example 2, continued}
\end{flushleft}

We use the same trait $x \in \Omega$ to characterize males and females.  The trait $x$ may represent promiscuity.  We start from \eqref{5.2} but now $F$ is a 2-vector and $A$ a $2 \times 2$-matrix with zero's on the diagonal. We assume
\begin{equation}\label{8.18}
A_{ij}(\tau,x,\xi)=a_i(x)b_j(\tau)c_j(\xi).
\end{equation}
Let
\begin{equation}\label{8.19}
\int_{-\infty}^{t}F_i(\sigma,x)d\sigma=a_i(x)w_i(t),
\end{equation}
then
\begin{equation}\label{8.20}
\begin{aligned}
w_1(t)&=N_2\int_{0}^{\infty}b_2(\tau)\int_{\Omega} c_2(\xi)(1-e^{-a_2(\xi)w_2(t-\tau)})\Phi_2(d\xi)d\tau, \cr
w_2(t)&=N_1\int_{0}^{\infty}b_1(\tau)\int_{\Omega} c_1(\xi)(1-e^{-a_1(\xi)w_1(t-\tau)})\Phi_1(d\xi)d\tau, \cr
\end{aligned}
\end{equation}
which is of the form \eqref{8.13} with $A=\begin{pmatrix} 0 & b_2 \cr b_1 & 0 \end{pmatrix}$ and
\begin{equation}\label{8.21}
\Psi(w)=\begin{pmatrix} N_1\int_{\Omega} c_1(\xi)(1-e^{-a_1(\xi)w_1(t-\tau)})\Phi_1(d\xi) \cr
N_2\int_{\Omega} c_2(\xi)(1-e^{-a_2(\xi)w_2(t-\tau)})\Phi_2(d\xi) \cr
\end{pmatrix}.
\end{equation}
Now we assume that \eqref{8.16} holds.  Then \eqref{8.17} holds, with now $\Psi$ as defined above.

 In our presentation of this example we avoided to discuss the consistency requirement that there are, in total, as many contacts of males with females as there are contacts of females with males. Partly we did so because transmission risk may be asymmetric, which obviously impairs the symmetry requirement. But the more important reason is that our aim here is just to illustrate the flexibility of the bookkeeping framework (and {\it not} to analyse a model of a heterosexually transmitted disease).

\section{Concluding remarks}\label{final}

 On November 13, 2022, the KM paper \cite{kermack1927} had, according to Google Scholar, 12.590 citations. On that same date the Royal Society listed 60.773 downloads since the beginning of 1997, when the paper became available online. No doubt there are among the authors that cite \cite{kermack1927} some who actually read the paper and who refer to the general age-of-infection model, see e.g. \cite{brauer2005, breda2012, diekmann2012, inaba2017, montalban2022, novozhilov2008, novozhilov2012, thieme2003, tkachenko2021}. In the big majority of cases, however, it is explicitly stated that \cite{kermack1927} introduced the SIR compartmental model and implicitly suggested that that is it. This both reflects and reinforces an incessant misconception in the math-epi community at large, viz., that \cite{kermack1927} is just about the SIR compartmental model.

   In fact, as shown in \cite{diekmann2018}, any compartmental model in which the (probability distribution of the)  state-at-infection is described by a given fixed vector $V$ corresponds to a (very) special case of the renewal equation. The compartmental system is coded by the triple $(V, \Sigma, U)$, with the vector $V$ describing the initial state, the matrix $\Sigma$ specifying the rates at which state transitions occur and the vector $U$ describing the contribution of the various states to the force of infection. (When there are several possibilities for the state-at-infection, one needs a system of renewal equations and a corresponding number of vectors $V$ and $U$, see \cite{diekmann2018} and Example 8.2.)

   It has long been recognized that host heterogeneity can have a big impact on epidemic dynamics, see, e.g., \cite{diekmann1990, diekmann2012, inaba2017, katriel2012, novozhilov2008, novozhilov2012, veliov2016, tsachev2017, hickson2014}. In general, the incorporation of heterogeneity necessitates the introduction of kernels and leads to infinite dimensional dynamical systems. In this context too, the question arises whether or not one can reduce to a finite system of ODE. In \cite{novozhilov2008, novozhilov2012} and in the more recent Covid-triggered papers \cite{gomes2022, montalban2022, neipel2020, rose2021} a restricted form of static heterogeneity, in which the traits of the two individuals involved in a contact are assumed to have independent influence on the likelihood of contact and/or transmission, is considered. The impact of such heterogeneity on the outbreak dynamics, is, of course, most easily studied if the heterogeneity is captured by a modification of the compartmental model one is interested in.
   
   Here we have shown that it is straightforward to derive the desired modification if one first formulates the heterogeneous version of the KM RE and relates it to an integrated version of the compartmental model. Once the modified integrated version is available, it is easy to derive the modification of the standard form by differentiation. The end result only involves (the derivative of) a function $\Psi$ from $\mathbb R$ to $\mathbb R$ defined in terms of the distribution of the trait and the functions that describe how susceptibility and infectiousness depend on the trait (so one can use the end result without any reference to renewal equations).
   
   From a more general theoretical point of view, our methodology is in the spirit of \cite{diekmann2020, diekmann2020b} and the much older references in there. In a similar manner, \cite{novozhilov2008} and \cite{novozhilov2012} build on the ideas of G.P. Karev as described in \cite{karev2010, karev2019} and the much older references in there.

   In \cite{breda2012}, it is shown how to extend the RE formulation to models incorporating demographic turnover and/or temporary immunity. Such models allow for endemic steady states. They do {\it not} have an integrated version, for the simple reason that the integrals are bound to diverge. At present it remains unclear whether or not one can include separable static heterogeneity in such models via a simple modification (the authors are not very optimistic ...).
   
   We refer to \cite{tkachenko2021} for an up to date Covid inspired account of the role of heterogeneity in outbreak dynamics. In there it is emphasized that, on top of persistent static heterogeneity, dynamic heterogeneity may play a major role in damping the overshoot of the herd immunity threshold (in much the same way as a prey refuge dampens prey-predator oscillations). An additional variable $h$ is introduced to capture the dynamic heterogeneity.
   
   Our message in this paper is a bit equivocal. On the one hand, our aim was to show how separable static heterogeneity can be easily incorporated into compartmental models. On the other hand, we want to emphasize that the RE formulation is far more general and flexible and that the predominance of compartmental models is rather detrimental for epidemic modeling. No doubt the fact, that user friendly tools for the numerical study of RE are lacking, contributes to their unpopularity. In order to end at a positive note, we call attention to two recent developments: i) discrete time models, see \cite{diekmann2021, kreck2022, sofonea2021} and ii) pseudospectral approximation, see \cite{scarabel2021}.

\backmatter


\bmhead{Acknowledgments}
We thank Toshikazu Kuniya for his technical assistance with the production of this manuscript, and
we also thank a referee for both constructive criticism and calling our attention to some relevant references that we had missed.

\section*{Declarations}
 `Not applicable' for that section.

\begin{appendices}






\end{appendices}


\bibliography{sn-bibliography}


\end{document}